\documentclass[amsmath,twocolumn,superscriptaddress,showpacs,aps,prb]{revtex4} 
\usepackage{bm}
\usepackage{graphics}
\begin{document}

\title{ Full counting statistics of a charge pump in the
  Coulomb blockade regime } 

\author{A.~V.~Andreev}
\affiliation{Bell Labs, Lucent Technologies, 600 Mountain Ave, Murray Hill, NJ 07974} 
\affiliation{Department of Physics, University of
  Colorado, CB 390, Colorado 80309-0390} 
\author{E.~G.~Mishchenko }
\affiliation{Bell Labs, Lucent Technologies, 600 Mountain Ave, Murray Hill, NJ 07974} 
\affiliation{Department of Physics, University of
  Colorado, CB 390, Colorado 80309-0390} 
\affiliation{L.~D.~Landau
  Institute for Theoretical Physics, Russian Academy of Sciences, 2 Kosygin St., Moscow 117334, Russia}

\begin{abstract}
  We study full charge counting statistics (FCCS) of a charge pump based
  on a nearly open single electron transistor.  The problem is mapped onto
  an exactly soluble problem of a non-equilibrium $g=1/2$ Luttinger liquid
  with an impurity. We obtain an analytic expression for the generating
  function of the transmitted charge for an arbitrary pumping strength.
  Although this model contains fractionally charged excitations only {\em
    integer} transmitted charges can be observed. In the weak pumping
  limit FCCS correspond to a Poissonian transmission of particles with
  charge $e^*=e/2$ from which all events with odd numbers of transferred
  particles are excluded.
\end{abstract}
\pacs{PACS numbers: 73.23.Hk, 73.40.Gk, 72.10.Bg, 71.10.Pm}
\maketitle

  Charge pumping has attracted considerable theoretical and experimental
  interest.  It occurs when the Hamiltonian of the system changes slowly
  in comparison with the characteristic time scales of the problem.  At
  the end of the pumping cycle, when the Hamiltonian returns to its
  initial value, a finite charge may be transmitted through the system.
  The amount of the transferred charge depends on the details of the
  pumping cycle.  This idea is due to Thouless,\cite{Thouless83} who
  showed that in certain one-dimensional systems the transmitted charge is
  quantized in the adiabatic limit.  Most of research efforts have
  focussed on charge pumping through mesoscopic devices.~\cite{Niu,Beal-Monod,Kouwenhoven,Brouwer98,Shutenko,Marcus,Aleiner98,Keller98,Andreev00,Simon00,Chamon00,Levitov01}

  Motivated by the efforts to build an accurate standard of electric
  current most experiments concentrated on single electron pumps in which
  the charge pumped during one cycle is quantized due to the Coulomb
  blockade effects.~\cite{Kouwenhoven,Keller98} Such devices are 
 already used in metrological applications to produce an accurate
  capacitance standard.~\cite{Keller98} 
  
  Understanding of noise properties of the pumped current and of the
  accuracy of quantization of the pumped charge are very important for
  metrological applications.  In this case it is desirable to know not
  only the average pumping current and its second moment (noise power)
  but the whole distribution function of the pumped charge.  Such full
  charge counting statistics (FCCS) were first considered in
  Refs.~\onlinecite{Levitov,coherent} for systems with non-interacting
  electrons. 
  
\begin{figure}
\resizebox{.38\textwidth}{!}{\includegraphics{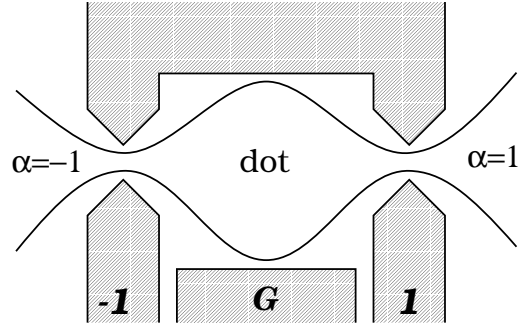}}
\caption{\label{fig:1} Schematic drawing of a single electron
  transistor electrostatically defined on a surface of a two dimensional
  electron gas.  The quantum dot is connected to two leads by single
  channel quantum point contacts labeled by $\alpha$.  The voltages on the
  gates $G$ and $\pm1$ determine respectively the average electron number
  in the dot, $N(t)$ and the reflection amplitudes, $r_{\pm 1}(t)$ in
  QPCs.}
\end{figure}

  In the present paper FCCS for a charge pump based on a
  single electron transistor are considered.  More precisely, the
  device in question consists of a quantum dot connected to the left and
  right leads by single channel quantum point contacts labeled by the
  index $\alpha=\pm1$, see Fig.~\ref{fig:1}.  Such devices can be
  fabricated in semiconductor heterostructures~\cite{review} where the
  electrons in the two-dimensional electron gas (2DEG) in a
  heterostrocture are electrostatically confined to the area of the dot by
  a negative voltage which is applied to the metallic gates located on top
  of the 2DEG.  The reflection amplitudes $r_\alpha$ in the contacts are
  controlled by the voltages on gates $\alpha$ and are assumed to be small
  throughout the pumping cycle, $r_\alpha \ll 1$.  The Coulomb interaction
  of electrons in the dot can be treated within the constant interaction
  model:
\begin{equation}
\label{eq:cim}
H_C=E_C(\hat{N}-N(t))^2,
\end{equation}
where $\hat N$ is the number of electrons in the dot, $E_C$ is the
charging energy and $N(t)$ is the dimensionless parameter proportional to
the voltage on the central gate $G$. 

At low temperatures, $T\ll E_c$, the electron transport across the device
is dominated by cotunneling processes.  The quantum dot
is assumed to be sufficiently large so that elastic cotunneling
effects~\cite{Aleiner97} can be neglected and the transport of electrons
across the device is dominated by the inelasting cotunneling.~\cite{Matveev95,Furusaki95} 
In addition the electrons are assumed to be
spin-polarized.  This can be realized experimentally by applying a strong
magnetic field parallel to the plane of the 2DEG.  

Sufficiently strong pumping can lead to a non-equilibrium distribution of
electrons in the dot. Below we assume that the deviations from equilibrium
may be neglected. This requirement imposes a limitation on the number of
pumping cycles in the absence of energy relaxation in the dot. Indeed,
inelastic cotunneling can be thought of as a coherent process in which an
electron, say from the left lead, enters a certain quantum state in the
dot and an electron from a different state in the dot leaves into the
right lead. As a result of such a process an electron is transferred
across the device and an electron-hole pair is created in the dot.  Upon
completion of $N$ pumping cycles roughly $N$ electron-hole pairs will be
created.  The number of electron-hole pairs in equilibrium may be
estimated as $T/\delta_1$, where $\delta_1$ is the single particle mean
level spacing in the dot.  Therefore, for the deviations from equilibrium
to be small we assume that the dot is sufficiently large so that the mean
level spacing $\delta_1\ll T/N$. In the presence of energy relaxation in
the dot this condition may be relaxed.

At frequencies below the charging energy $E_C$ the pumping cycle 
is described by a single complex variable,~\cite{Furusaki95,Aleiner98}
$z(t)=r_1(t)\exp(i\pi N(t))+r_{-1}(t)\exp(-i\pi N(t))$. The average
pumping current for this cycle was obtained in Ref.~\onlinecite{Aleiner98}.

Here we study FCCS for this pump.  The probability distribution function
$P_N(Q)$ for the charge $Q$ transmitted through the dot upon completion of
$N$ pumping cycles is determined by the generating function
  \begin{equation}
    \label{eq:genfuncdef}
   F_N(\lambda)= 
  \sum\limits_{Q}\exp(i\lambda Q)P_N(Q),
  \end{equation}
  where the charge $Q$ is measured in units of the absolute value of the
  electron charge, $e$, and the sum goes over all its possible values. The
  $n$-th cumulant $\langle \langle Q^n \rangle \rangle$ of the transmitted
  charge may be determined from $F_N(\lambda)$ through the relation
  \begin{equation}
    \label{eq:cumulants}
\langle
  \langle Q^n \rangle \rangle=\left. \frac{d^n \ln F_N(\lambda)}{i^n
      d\lambda^n}\right| _{\lambda=0}. 
  \end{equation}

  Below we concentrate on the pumping cycle in which $z(t)=z_0
  \exp(-i\omega t)$.  In this case we obtain the following generating
  function
\begin{eqnarray}
\label{eq:result}
&&\ln F_N(\lambda)=\frac{1}{2}
\sum\limits_{l=-\infty}^\infty \ln\left(1+\cos^2\theta(\varepsilon_l)\left
[ n_-(\varepsilon_l) 
[1-n_+(\varepsilon_l)]
\right. \right. \nonumber \\ 
&&\times \left. \left.  \left(e^{i\lambda}-1\right) + n_+(\varepsilon_l)
[1-n_-(\varepsilon_l)]\left(e^{-i\lambda}-1\right)\right]\right) -iN\lambda .
\end{eqnarray}
Here $\varepsilon_l=\omega(2l+1)/(2N)$, with $l$ being an integer, denotes
the discrete fermionic frequency, $n_\pm(\epsilon)=n_0(\epsilon\pm
\omega)=(e^{(\epsilon+\omega)/T}+1)^{-1}$ is the Fermi distribution
function shifted by $\pm \omega$, and $\exp[i\theta(\epsilon)]=(\epsilon
+i\Gamma)/\sqrt{\epsilon^2 + \Gamma^2}$, where $\Gamma=2\gamma |z_0|^2
E_C/\pi^2$, $\gamma=\exp C$, with $\ln\gamma = C\approx 0.5772\ldots$
being the Euler constant.

Note that since the generating function $F_N(\lambda)$ in
Eq.~(\ref{eq:result}) is periodic in $\lambda$ with the period $2 \pi$,
$F_N(\lambda +2 \pi) = F_N(\lambda)$, only {\em integer} values of charge
$Q$ can be transmitted.

Eq.~(\ref{eq:result}) acquires a particularly simple form at low
temperatures, $T\ll \omega$.  Approximating the Fermi functions by the
step functions $n_{\pm}(\epsilon_l)=\Theta(-\epsilon_l \mp \omega)$ we
observe from Eq.~(\ref{eq:result}) that only the energy interval $-\omega
< \epsilon_l < \omega$ contributes to the pumped charge. Using the Poisson
summation formula we can write Eq.~(\ref{eq:result}) as
\begin{eqnarray}
\label{poisson}
\ln F_N(\lambda)& =& -iN\lambda +
\Upsilon N \sum\limits_{n=-\infty}^\infty 
\int\limits_{-1/\Upsilon}^{1/\Upsilon} d x ~
e^{i n (x \Gamma \tau+\pi)} \nonumber\\
&& \times
\ln{x^2 e^{i\lambda/2}+1}-\ln{x^2 +1},
\end{eqnarray}
where $\Upsilon=\Gamma/\omega$ is the relative pumping strength.  For long
observation times $\tau\Gamma =2\pi \Upsilon N \gg 1$ the terms $n\neq 0$
become small due to the presence of quickly oscillating factors in their
integrands.  Explicit evaluation of the main, $n=0$, term gives
\begin{eqnarray}
\label{generator}
\ln{ F_N(\lambda)}&=& N \ln{\frac{\Upsilon^2 e^{-i\lambda}+1}
{\Upsilon^2 +1}}-2\Upsilon N 
\arctan{(\Upsilon^{-1})}
\nonumber\\
&&+2  \Upsilon N e^{-i\lambda/2}
\arctan{\left(\Upsilon^{-1} e^{i\lambda/2} \right)}.
\end{eqnarray}
With the aid of Eq.~(\ref{eq:cumulants}) we obtain for the average
pumping current,
\begin{equation}
\label{mean}
I=\frac{e\langle Q \rangle}{\tau} = -\frac{e
\Gamma}{2\pi} 
\arctan{(\Upsilon^{-1})}.
\end{equation}
The initial growth of the current with the pumping frequency $\omega$
saturates at $I=- e\Gamma/4$ for large $\omega$.

In the strong pumping limit, $\Upsilon \gg 1$, Eq.~(\ref{generator})
yields
\begin{equation}
\label{strong}
\ln{ F_N(\lambda)}=-iN\lambda + \frac{N}{3 \Upsilon^2}
\left(e^{i\lambda}-1 \right),
\end{equation}
where the first term contributes only to the average current, and the
second term describes a Poisson process for particles with an integer
charge $e^*= e$ and transmission frequency $\omega/(6\pi \Upsilon^2)$.

In the limit of weak pumping, $\Upsilon \ll 1$, we can perform the
integration over $x$ in Eq.~(\ref{poisson}) from $-\infty$ to $\infty$.
Retaining the terms with $n \neq 0$ we obtain
\begin{equation}
\label{gen_fun}
F_N(\lambda)=\frac{\cosh{[\pi \Upsilon N
e^{-i\lambda/2}]}}{\cosh{[\pi\Upsilon N]}}.
\end{equation}
To evaluate the cumulants (\ref{eq:cumulants}) it suffices to know
$F_N(\lambda)$ at $\lambda \to 0$. For long observation times $\tau \Gamma
\gg 1$ we can neglect the exponentially small terms in Eq.~(\ref{gen_fun})
and write the logarithm of the generating function as,
\begin{equation}
\label{weak}
\ln{ F_N(\lambda)}=\pi\Upsilon N
\left(e^{-i\lambda/2}-1 \right).
\end{equation}
This formula can also be derived directly from Eq.~(\ref{generator}). It
corresponds to a Poisson process which describes independent transmission
of quasiparticles with the average transmission frequency $\Gamma/2$ and
{\em fractional} charge $e^*=e/2$.  The true limiting expression for weak
pumping, Eq.~(\ref{gen_fun}) is periodic in $\lambda$ with the period
$2\pi$, allowing transmission of only integer charges. It is easy to check
that Eq.~(\ref{gen_fun}) describes a Poisson process for charge $e/2$
particles from which all transmission events with odd numbers of
transferred particles have been excluded.  The corrections to
Eq.~(\ref{gen_fun}) are small $\sim \Upsilon^2$.

One may define the effective charge $e^*$ of the carriers through the
ratio of the variance of the transmitted charge to its average value for
intermediate pumping strengths as well. However
Eqs.~(\ref{weak},\ref{strong}) show it can be interpreted as a charge of
independently transmitted particles only in the limits of weak, $\Upsilon
\ll 1$, and strong, $\Upsilon \gg 1$, pumping.  The coefficients in the
Taylor expansion of Eq.~(\ref{generator}) in powers $1/\Upsilon^n$ (for
strong pumping) or in powers $\Upsilon^n$ (for weak pumping) represent
Poissonian transmission processes of multiple charge $ne^*$, in agreement
with Ref.~\onlinecite{Saleur00}.

Below we present the derivation of the above results. At $T, \omega \ll
E_C$ the pumping cycle is described by the
Hamiltonian~\cite{Matveev95,Aleiner98},
\begin{eqnarray}
\label{ham}
H&&=\int_{-\infty}^{\infty} dk \left[\frac{\epsilon_k}{2} \Psi_k^{\dagger}
  \sigma_3 \Psi_k 
  +\frac{\kappa \zeta \Psi_k^{\dagger}}{\sqrt{2\pi}} 
\left( \begin{array}{c} -z^*(t)\\ z(t)
    \end{array} \right)   
 \right],
\end{eqnarray}
where $\kappa = \sqrt{\gamma v E_C/\pi^2}$. In Eq.~(\ref{ham}) $\Psi_k $
is a vector fermion operator in Gorkov-Nambu notations and is expressed
through the creation and annihilation operators $c_k$ and $c^\dagger_{k}$
as $\Psi_k^{\dagger} = (c_k^\dagger, c_{-k} )$, and $\sigma_3$ is the
Pauli matrix.  In this model electrons have a linear spectrum
$\epsilon_k=v k$ and are coupled to a resonant state described by a
Majorana fermion $\zeta$, $\zeta^2=1$.  The current through the pump is
given by
\begin{equation}
 I=-\frac{e v_F}{2}  \int_{-\infty}^{\infty} dk ~\Psi^\dagger_k  \Psi_k  .
\end{equation}

For the pumping cycle considered here the gauge transformation 
\begin{equation}
\label{gauge}
\Psi_k \to \exp(i\sigma_3\omega t )\Psi_k ,
\end{equation}
removes the time dependence of the Hamiltonian, $z(t) \to z_0$.  As a
result, the chemical potentials of electrons and holes shift by $\pm
\omega$.  The current operator in this gauge acquires an additional
anomalous term and takes the form,
\begin{eqnarray}
I = -\frac{e v_F}{2} \int_{-\infty}^{\infty} dk
~\Psi^\dagger_k  \Psi_k +\frac{e\omega}{2\pi}.
\label{currentgauge}
\end{eqnarray}

The stationary Hamiltonian (\ref{ham}) (with $z(t)\to z_0$) was diagonalized 
by Matveev~\cite{Matveev95} in terms of the linear combinations of particle
and hole operators, 
\begin{subequations}
\label{operators}
\begin{eqnarray}
\widetilde{C_k}&&=\frac{c_k+c_{-k}^{\dagger}}
{\sqrt{2}},\\
C_k&&= \frac{\epsilon_k \pm
  i\Gamma}{\sqrt{\epsilon_k^2+\Gamma^2}}\frac{c_k-c_{-k}^{\dagger}} 
{\sqrt{2}}- \zeta
\sqrt{\frac{v_F \Gamma}{2\pi (\epsilon_k^2+\Gamma^2)}}\nonumber\\ 
&&\mbox{}+\frac{\Gamma}{\pi \sqrt{\epsilon_k^2+\Gamma^2}}
\int \frac{d\epsilon_{k'}}{\epsilon_k -\epsilon_{k'} \pm i0}
\frac{c_{k'}-c^{\dagger}_{-k'}}{\sqrt{2}}.
\label{C}
\end{eqnarray}
\end{subequations}
Both signs in Eq.~(\ref{C}) give equivalent expressions.  For the
upper/lower sign the last term in Eq.~(\ref{C}) gives a vanishing
contribution to $\Psi(x)$ at $x \to \pm \infty$ after a Fourier
transformation to the real space.  (The second term in Eq.~(\ref{C})
corresponds to the resonant state vanishing for $x \to \pm \infty$.)

Having observed these asymptotic properties of operators (\ref{operators})
we can readily build scattering states corresponding to the scattering of
an electron,
\begin{eqnarray}
\label{electron}
&&\frac{1}{\sqrt{2}} \left[e^{-i\theta(\epsilon_k)}\widetilde{C_k}+C_k
\right] \nonumber\\ 
&&=\left\{\begin{array}{ll}
e^{-i\theta(\epsilon_k)}c_k, & x \to -\infty, \\
  \left[c_k \cos{\theta(\epsilon_k)}
-i c_{-k}^{\dagger}\sin{\theta(\epsilon_k)}\right], & x \to +\infty, 
\end{array}\right.
\end{eqnarray}
and a hole
\begin{eqnarray}
\label{hole}
&&\frac{1}{\sqrt{2}} \left[e^{-i\theta(\epsilon_k)}\widetilde{C_k}-C_k
\right]\nonumber\\ 
&&= \left\{ \begin{array}{ll}
e^{-i\theta(\epsilon_k)}
c_{-k}^{\dagger}, & 
x \to -\infty,    \\
 \left[-i c_{k}\sin{\theta(\epsilon_k)}
+ c_{k}^{\dagger} \cos{\theta(\epsilon_k)}\right], & x \to +\infty, 
\end{array}\right.
\end{eqnarray}
where $\cos{\theta(\epsilon_k)}=\epsilon_k/\sqrt{\epsilon_k^2+\Gamma^2}$,
and $\sin{\theta(\epsilon_k)}=\Gamma/\sqrt{\epsilon_k^2+\Gamma^2}$.
We can now write the scattering matrix 
\begin{equation}
\label{matrix}
\hat{S}(\epsilon_k)=e^{i\theta(\epsilon_k)} \left(
\begin{array}{cc}
 \cos{\theta(\epsilon_k)} & -i \sin{\theta(\epsilon_k)} \\
-i \sin{\theta(\epsilon_k)} & \cos{\theta(\epsilon_k)}
\end{array}
\right),
\end{equation}
for the scattering between electron (positive current) 
and hole (negative current) states. 

Thus, the problem reduces to the problem of non-equilibrium chiral
fermions scattering off a resonant state at zero energy. The electrons and
holes here are characterized by non-equilibrium distribution functions
$n_\pm(\epsilon)$ defined below Eq.~(\ref{eq:result}) and may be
represented by the diagonal matrix
\begin{equation}
  \label{eq:nhat}
\hat{n}(\epsilon)= \left(
\begin{array}[c]{cc}
n_-(\epsilon)& 0 \\
0& n_+(\epsilon)
\end{array}\right) .  
\end{equation}

The full counting statistics for non-equilibrium non-interacting fermions
have been extensively studied.~\cite{Levitov} The generating function of 
the transmitted charge is given by
\begin{eqnarray}
  \label{eq:genfunc}
 && F_N(\lambda)= \exp(-iN\lambda) \times \nonumber \\
&&  \exp \left\{  {\rm Tr}\sum\limits_{k=0}^\infty
  \ln \left[ \openone + \hat{n}(\epsilon_k)\left(
      \hat{S}^\dagger_{-\lambda} (\epsilon_k)\hat{S}_\lambda (\epsilon_k)- 
\openone\right)\right]\right\},
\end{eqnarray}
where $\hat{S}_{\pm \lambda} (\epsilon_k)= \exp(\pm \frac{i}{4}\sigma_3
\lambda) \hat{S} (\epsilon_k) \exp(\pm \frac{i}{4}\sigma_3 \lambda)$ with
$\hat{S} (\epsilon_k)$ defined in Eq.~(\ref{matrix}).  The first term in
this equation arises from the anomalous term in the current operator,
Eq.~(\ref{currentgauge}). Since electron and hole operators describe the
same physical states, the sum over energies is restricted to positive
frequencies in order to avoid double counting of degrees of freedom.
Substituting Eqs.~(\ref{matrix}) and (\ref{eq:nhat}) into
Eq.~(\ref{eq:genfunc}) we obtain the generating function
(\ref{eq:result}).

In conclusion, we have obtained full counting statistics for a charge pump
based on a nearly open single electron transistor.  In the spin-polarized
case the problem is mapped onto an exactly soluble chiral fermion model,
Eq.~(\ref{ham}).  In the weak pumping regime $\Gamma \ll \omega$ the
generating function given by Eq.~(\ref{gen_fun}) corresponds to a Poisson
process of charge $e^*=e/2$ particles with transmission rate $\Gamma/2$
from which all events with odd numbers of transferred particles are
excluded.  Alhough all the moments of the transferred charge obtained from
Eq.~(\ref{gen_fun}) are practically indistinguishable from those of a
simple Poisson process for charge $e^*=e/2$ particles only integer
transferred charges may be observed in a pumping experiment. Since the
Hamiltonian (\ref{ham}) of this model describes a $g=\frac{1}{2}$
Luttinger liquid with an impurity one may expect that similar
conclusion hold for other coupling strengths $g \neq \frac{1}{2}$ and
other problems with fractionally charged excitations realized for example
in Quantum Hall experiments. \cite{Samin97,Reznikov97,Simon00}  

Equation (\ref{gen_fun}) differs from the weak pumping result in
Ref.~\onlinecite{Levitov01}.  The reason for this discrepancy and for charge
fractionalization lies in the failure of perturbation theory in the
reflection amplitudes $r_{\pm 1}$ for our model at sufficiently
low energies $\epsilon < \Gamma $.  The effective reflection
coefficient $\Gamma^2/(\epsilon^2+\Gamma^2)$ that determines the strength
of pumping approaches unity in this energy range.  Thus, the true
expansion parameter at weak pumping is not the reflection amplitude
$r_{\pm 1}$ but the ratio of energy scales $\Upsilon =\Gamma/\omega$.
Perturbation theory in the reflection amplitude fails for Luttinger models
with an impurity at other interaction strengths $g \neq \frac{1}{2}$ as well
wich leads to the appearance of an energy scale analogous to $\Gamma$
below which the system is in the strong coupling limit.\cite{Kane94,Saleur00}

Although we have focused on the low temperature case $T \ll \omega$ the
validity of the result (\ref{eq:result}) is restricted only by the
condition $T < E_C$.  Using Eq.~(\ref{eq:result}) we obtain the general
expression for the average pumping current,
  \begin{equation}
    \label{eq:pumpcurrent}
    I=-\frac{e\omega}{2\pi}+\frac{e}{4\pi}
 \int_{-\infty}^{\infty}
   \frac{\epsilon^2 d\epsilon }{\epsilon^2+\Gamma^2}~
\frac{\sinh{\frac{\omega}{T}}}{ \cosh{\frac{\epsilon}{T}}+
 \cosh{\frac{\omega}{T}}}.
\end{equation}
This formula reduces to the result 
of Ref.~\onlinecite{Aleiner98} in the linear response
regime $\omega \ll T$.

The case of zero pumping and finite external bias $V$ can be obtained from
the above expressions by substituting $\omega \to eV$ and omitting the
anomalous term in the current.  For example the non-linear $I-V$
characteristic is obtained in this way from the second term of Eq.~(\ref{eq:pumpcurrent}).

We are grateful to I.~Aleiner, B.~Altshuler, L.~Glazman, A.~Kamenev,
L.~Radzihovsky and S.~Sondhi for valuable discussions.  This research was
sponsored by the Grants DMR-9984002 and BSF-9800338 and by the A.P.~Sloan
and the Packard Foundations.  A.~A.~gratefully acknowledges the warm
hospitality of the Centre for Advanced Studies in Oslo.  E.~M.~thanks
the Russian Foundation for Basic Research, Grant 01-02-16211.

\end{document}